\documentclass[dvips]{acta}
\usepackage{supertabular,lscape,epsfig}
\usepackage{amssymb}
\usepackage{amsmath}

\SetPages{0}{0}

\SetVol{58}{2008}

\begin{document}

\begin{Titlepage}
\Title{\bf Spectroscopic Study of Candidates for Kepler Asteroseismic Targets -- \\
Solar-Like Stars
\footnote{The data used in this paper have been obtained at the {\it
M.G. Fracastoro\/} station of the Catania Astrophysical Observatory
and the F.L.\ Whipple Observatory, Mount Hopkins, Arizona.}
}

\Author{J.~M~o~l~e~n~d~a~-~\.Z~a~k~o~w~i~c~z$^1$, A.~F~r~a~s~c~a$^2$ and D.W.~L~a~t~h~a~m$^3$
}
{$^1$Astronomical Institute, University of  Wroc{\l }aw, Kopernika 11, Wroc{\l }aw, Poland\\
e-mail: molenda@astro.uni.wroc.pl\\
$^2$Catania Astrophysical Observatory, Via S.Sofia 78, Catania, Italy\\
e-mail: afr@oact.inaf.it\\
$^3$Harvard-Smithsonian Center for Astrophysics, 60 Garden Street, Cambridge, MA 02138, USA\\
e-mail: dlatham@cfa.harvard.edu}

\Received{Month Day, Year}
\end{Titlepage}

\Abstract{ We report spectroscopic observations of 23 candidates for Kepler 
asteroseismic targets and 10 other stars in the Kepler field, carried out at two 
observatories (see the footnote). For all these stars, we derive the radial 
velocities, effective temperature, surface gravity, metalicity, 
the projected rotational velocity, and estimate the MK type.

HIP\,97513 and HIP\,92132 are classified as suspected new single-lined spectroscopic 
binaries. 

For 28 stars, the radial velocity is measured for the first time.}{Space missions: 
Kepler -- Stars : radial velocities -- Binaries: spectroscopic -- Stars: atmospheric parameters} 

\section{Introduction}

Kepler\footnote{http://kepler.nasa.gov/}, a NASA space mission, shall be launched 
in April 2009 with the goal of detecting Earth-size and larger planets by 
means of the method of photometric transits (Borucki et al.~1997). 
Kepler photometry will also be used for detecting pulsations in program stars 
and deriving accurate values of their pulsation frequencies. This will allow an 
investigation of internal structure of these stars by means of asteroseismology 
(see, e.g., Christensen-Dalsgaard 2004). 

In this paper, which is a sequel to the spectroscopic study of candidates for Kepler 
asteroseismic targets published by Molenda-\.Zakowicz et al.~(2007, henceforth Paper I),
we discuss 23 candidates for asteroseismic targets for Kepler and 10 other stars. 
Unfortunately, all 33 stars have Hipparcos parallaxes not precise enough to compute the 
luminosities, and therefore they were listed in Molenda-\.Zakowicz et 
al.~(2006) as secondary candidates for Kepler asteroseismic targets, SATS. Since 
the stars have spectral types F, G or K, they are expected to show solar-like 
oscillations with amplitudes detectable in Kepler photometry. 

Ten of these stars fall either just beyond the Kepler CCD chips or into star tracker 
corners, so that they are not expected to be observed in normal conditions. Therefore,
in the present paper we have narrowed down the definition of the SATS to the 23 stars that 
fall onto active chips of the Kepler CCDs. 

The paper is organised as follows.
After giving an account of the spectroscopic observations and reductions in Sect.\ 2, in 
Sect.\ 3 we discuss stars showing variable radial velocity. In Sect.\ 4, we determine the 
effective temperature, surface gravity, metalicity, and MK spectral type of the stars.
In Sect.\ 5, we give their projected rotational velocity. Sect.\ 6 contains a summary. 

\section{Observations and Reductions}

The observations were carried out at the {\it M.G. Fracastoro\/} station 
(Serra La Nave, Mount Etna, elevation 1750 m) of the Catania Astrophysical 
Observatory (CAO), Italy (39 spectrograms) and at the F.L.\ Whipple 
Observatory (FLWO), Mount Hopkins, Arizona (6 spectrograms).

At CAO, we used a 91-cm telescope and the fiber-fed echelle spectrograph 
FRESCO. The spectra were recorded with the resolving power R=21\,000 in a 
spectral range that covered about 2\,500 {\AA} in 19 orders. A thinned
back-illuminated CCD SITe chip (SI033B) with 1024x1024 24x24-$\mu$m pixels 
was used as the detector. At FLWO, we used the 1.5-m Tillinghast reflector 
and the CfA Digital Speedometer with the resolving power R=35\,000. 
An intensified photon-counting Reticon was used as the detector. In this system, 
a single 45 {\AA} spectrogram, centered at $\lambda \simeq$ 5187 {\AA}, was 
recorded in one exposure. 

The IRAF\footnote{IRAF is distributed by the National Optical Astronomy Observatories, 
which are operated by the Association of Universities for Research in Astronomy, 
Inc., under cooperative agreement with the National Science Foundation.} software 
was used for the reduction and calibration of the spectrograms measured at CAO.
For the spectrograms measured at FLWO, a special procedure described in 
detail in Latham et al.~(1992) was employed. Detailed description of the 
reduction of the data and the extraction of the spectra has been given in Paper I.

The radial velocities of the stars observed at CAO were determined by the
cross-correlation method provided by IRAF. For the templates, we used Arcturus or 
$\beta$ Oph for which precise values of R.V.\ are available (Udry et al.~1999) 
and which were observed on the same nights as the program stars. For the spectrograms 
measured at FLWO, a grid of synthetic spectra based on model atmospheres of R.L.\ 
Kurucz and computed by Jon Morse was used as the templates (see Torres et al.~2002). 

The majority of stars discussed in this paper had single spectrograms measured 
at CAO or FLWO. Nine stars were observed more than once; five were observed at 
both observatories. Therefore, before the analysis was started, the R.V.\ of 
these nine stars were moved to one reference system by subtracting 0.5 km\,s$^{-1}$ 
from each FRESCO's measurement. This offset has been calculated using the mean 
values of R.V.\ measured for 15 G-type dwarfs with FRESCO and the CfA Digital 
Speedometers, as described in Paper I. All results discussed in the present 
paper are obtained from the merged data.

\MakeTable{lcccrclc}{12.5cm}{Radial velocities (in km/s) of the program stars.
The code in the last column indicates stars that fall
onto the active chips of the Kepler CCDs, ``A'', stars that
fall into CCD gaps, ``G'', or stars that fall into star
tracker corners, ``S''. The individual velocities are not corrected for the shift
between observatories}
{\hline\noalign{\smallskip}
HIP    &$\alpha _{\rm 2000}$ &$\delta _{\rm 2000}$ &HJD -  &R.V. & s.e. & Instrument & code\\
       &                     &                     &2400000\\
\noalign{\smallskip}\hline\noalign{\smallskip}
 91841  & 18 43 28.30&+47 42 13.0&4266.5839 &   $-$18.70 &  0.15 &  FRESCO       & A\\
 91841  & 18 43 28.30&+47 42 13.0&4312.4190 &   $-$18.02 &  0.20 &  FRESCO       & A\\
 91841  & 18 43 28.30&+47 42 13.0&4363.3105 &   $-$17.50 &  0.15 &  FRESCO       & A\\
 91841  & 18 43 28.30&+47 42 13.0&4368.6002 &   $-$18.20 &  0.36 &  Tillinghast  & A\\
        &            &           &          &            &       &               & \\
 92053  & 18 45 44.79&+48 23 57.6&4266.3888 &   $-$35.84 &  0.28 &  FRESCO       & G\\
        &            &           &          &            &       &               & \\
 92132  & 18 46 41.64&+44 41 06.5&4279.5254 &    $-$3.57 &  0.27 &  FRESCO       & A\\
 92132  & 18 46 41.64&+44 41 06.5&4363.3650 &    $-$1.34 &  0.25 &  FRESCO       & A\\
        &            &           &          &            &       &               & \\
 92775  & 18 54 16.95&+42 59 00.4&4278.5515 &  $-271$.05 &  0.54 &  FRESCO       & A\\
        &            &           &          &            &       &               & \\
 92941  & 18 56 09.26&+46 39 56.5&4289.5676 &   $-17$.46 &  0.13 &  FRESCO       & G\\
        &            &           &          &            &       &               & \\
 92961  & 18 56 21.26&+45 30 53.1&4290.3441 &   $-$28.78 &  0.16 &  FRESCO       & A\\
 92961  & 18 56 21.26&+45 30 53.1&4313.5979 &   $-$28.54 &  0.33 &  FRESCO       & A\\
        &            &           &          &            &       &               & \\
 93320  & 19 00 27.13&+45 41 32.1&4268.4200 &   $-$47.47 &  0.53 &  FRESCO       & A\\
 93320  & 19 00 27.13&+45 41 32.1&4280.7682 &   $-$46.97 &  0.39 &  Tillinghast  & A\\
 93320  & 19 00 27.13&+45 41 32.1&4286.8822 &   $-$47.02 &  0.37 &  Tillinghast  & A\\
\multicolumn{8}{c}{\dotfill}\\
 98793  & 20 03 55.17&+44 08 24.2&4266.5154 &      13.99 &  0.37 &  FRESCO       & A\\
\noalign{\smallskip}\hline    
}

In Table 1, we give the individual radial velocity measurements not 
corrected for this offset. The table is available in electronic form 
from the Acta Astronomica Archive (see the cover page). A sample, 
containing the heading, the first 14 rows and the last row, is printed 
above. In the first column we give the HIP number, in the second and the third
column, the right ascension and declination, in the fourth, 
Heliocentric Julian Day of the middle of the exposure, in the fifth and 
sixth, the radial velocity, R.V., and the standard error, s.e., in the
next column, the instrument used, and in the last column, the information 
whether the star falls onto the active pixels of Kepler CCDs, coded with 
``A'', into the gaps between CCD chips, coded with ``G'', or into a star 
tracker corner, coded with ``S''. 

In Table 2, we list the nine stars for which we had two or more 
spectrograms. In the first column, we give the HIP number of the star, in 
the second, the number of spectrograms, in the third, the total time-span 
of observations in days, then, the mean radial velocity, R.V., 
in km\,s$^{-1}$, the ratio of external-to-internal error, e/i, the reduced 
$\chi ^2$, and the probability that a star with constant velocity will 
have $\chi ^2$ value larger than the observed one, $\rm P(\chi ^2)$. For a
detailed description of the method of computing the errors, we refer 
to Latham et al.~(2002).

\MakeTable{crrrcrrr}{12.5cm}{Nine stars for which more than one spectrogram 
was measured: the number of spectrograms measured, the total time-span
in days, the mean radial velocity, R.V., in km/s, the ratio of 
external-to-internal error, e/i, the reduced $\chi ^2$, and the probability 
that a star with constant velocity will have $\chi ^2$ value larger
than the observed one, $\rm P(\chi ^2)$}
{\hline\noalign{\smallskip}
HIP   & $N$ & span  & R.V. & s.e. & e/i & $\chi ^2$ & $\rm P(\chi ^2)$\\
\noalign{\smallskip}\hline\noalign{\smallskip}
 91841 & 4 & 102 & $-$18.09 & 0.29  &  2.71 & 0.32 & 0.96\\
 92132 & 2 &  84 &  $-$2.87 & 1.40  &  7.60 & 1.12 & 0.29\\ 
 92961 & 2 &  23 & $-$29.23 & 0.15  &  0.87 & 0.02 & 0.90\\ 
 93320 & 3 &  19 & $-$47.20 & 0.34  &  1.37 & 0.26 & 0.88\\ 
 93469 & 2 &  17 & $-$55.16 & 0.63  &  1.60 & 0.16 & 0.67\\ 
 94022 & 2 &  16 & $-$24.61 & 0.31  &  1.30 & 0.05 & 0.85\\
 94918 & 2 &  0.1& $-$47.01 & 0.58  &  0.80 & 0.33 & 0.57\\ 
 97513 & 2 &  59 &  $-$8.69 & 2.73  & 26.64 & 3.20 & 0.07\\
 98793 & 2 &  14 &    13.71 & 0.32  &  1.43 & 0.05 & 0.81\\ 
\noalign{\smallskip}\hline
}

\section{Program Stars with Variable Radial Velocity}

\subsection{HIP\,97513}

HIP\,97513 shows a very low value of $\rm P(\chi ^2)$. In the Hipparcos Catalogue 
(ESA 1997), it is is listed as a binary with an ambiguous double-star solution. 
HIP\,97513 was observed spectroscopically by Bartkevi\u{c}ius \& Sperauskas (2005) who 
measured the star's R.V.\ on two consecutive nights but found no variation. We 
have used their values to compute an improved weighted mean 
radial velocity, R.V.\ = $-$8.21 $\pm$1.06 km/s. 

We classify HIP\,97513 as a suspected single-lined spectroscopic binary.

\subsection{HIP\,92132}

HIP\,92132 is the second star from our sample which shows indications of 
variability in radial velocity. The R.V.\ measured by other observers range from 
$-$3.5$\pm$0.8 km/s (Gontcharov 2006), through $-$4.1 km/s (Wielen et al. 2000) to 
$-$4.2$\pm$19.9 (Bobylev et al. 2006). 

Since Wielen et al.~(2000) do not give the uncertainties of the R.V.\ and the 
uncertainty given by Bobylev et al.~(2006) is very high, we did not include these
data in calculating a mean radial velocity.

We classify HIP\,92132 as a suspected single-lined spectroscopic binary.

\section{Effective Temperature, Surface Gravity, Metalicity, and the MK Type}

\subsection{From a Comparison with Standard Stars}

We determined $T_{\rm eff}$, $\log g$  and $\rm [Fe/H]$ of the program stars 
observed with the FRESCO instrument, using ROTFIT, which is an IDL code developed 
by A.F.\ and his coworkers (see, e.g., Frasca et al.\ 2003, 2006). The method, 
originally developed by Katz et al.~(1998) and Soubiran et al.~(1998),
consists in comparing the spectra of program 
stars with a library of spectra of reference stars, and computing the weighted 
means of the astrophysical parameters of these five reference stars which best 
reproduce the target spectrum. The $T_{\rm eff}$, $\log g$, and $\rm [Fe/H]$ 
computed in this way are adopted as estimates of the astrophysical parameters of 
the program star. For the measure of the similarity of spectra, $\chi ^2$ is used.
This method, as discussed by, e.g., Frasca et al.~(2006), allows 
simultaneous and fast determination of $T_{\rm eff}$, $\log g$, and $\rm [Fe/H]$ 
even from spectrograms of low signal-to-noise ratio (S/N) or moderate resolution. 

\begin{figure}[htb]
\includegraphics{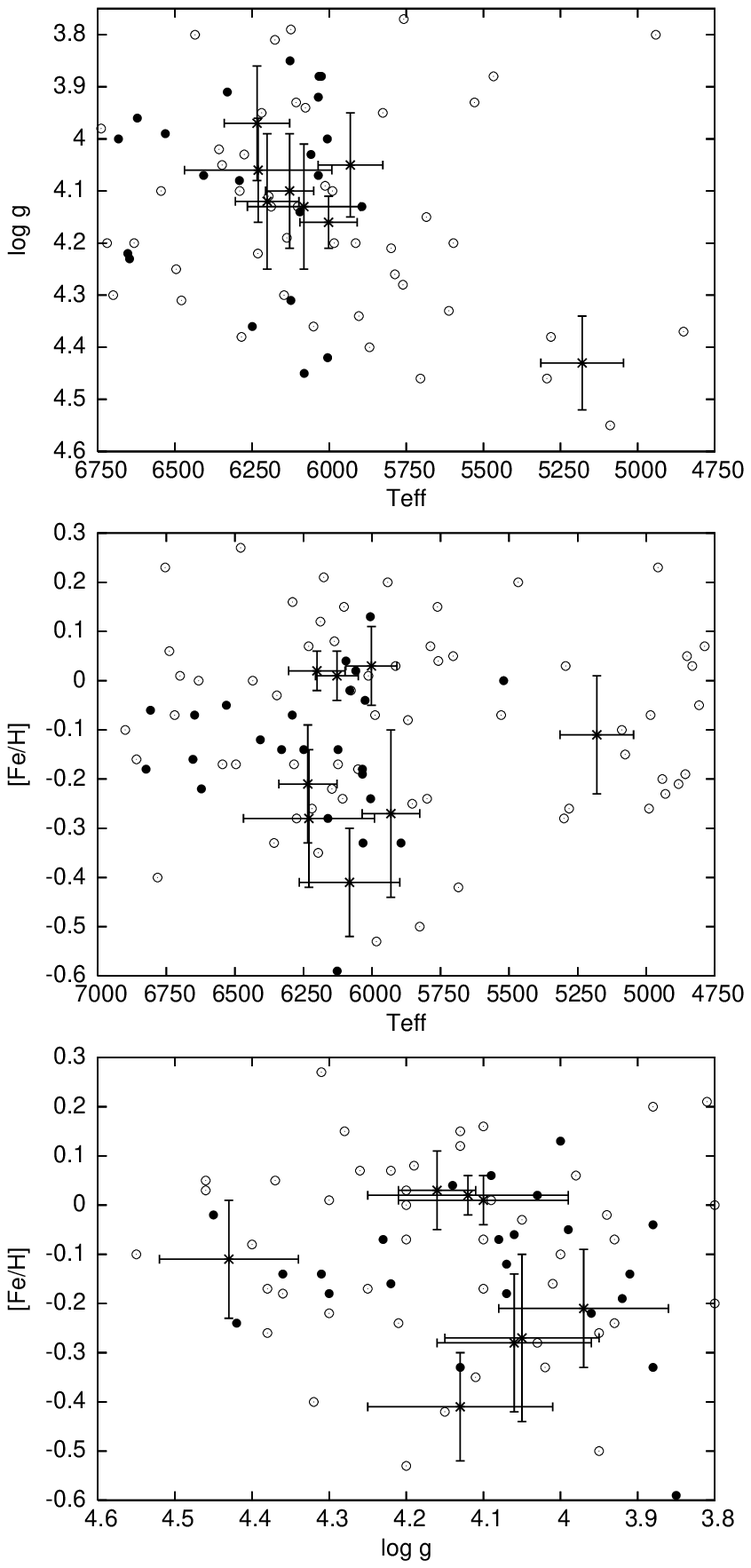}
\FigCap{Distribution of the FRESCO reference stars from Table 8 of Paper I (open circles), the
new reference stars added in this paper (dots) and the program stars from Table 3 (asterisks)
in the $T_{\rm eff}$--$\log g$, $T_{\rm eff}$--$\rm [Fe/H]$ and $\log g$--$\rm
[Fe/H]$ planes. For program stars, $1\sigma$ error bars are shown.
The diagrams show the regions occupied by the program stars.}
\end{figure}

For the reference stars, we used 240 slowly rotating stars of which 
spectrograms are available from the ELODIE archive (Prugniel \& Soubiran 2001).
These stars have been listed in Table 7 of Paper I. Then, similarly as in Paper I, we 
performed parallel computations with ROTFIT using 109 reference stars observed with 
FRESCO. The FRESCO grid consists of the 82 stars listed in Table 8 of Paper I and 27 new
reference stars which were observed in the same observing season as the program 
stars. In Table 5, which is available electronically from the Acta Astronomica Archive, we 
list all reference stars from the FRESCO library used in the present paper.

Adding the above-mentioned 27 new reference stars to the FRESCO grid did not change its 
limits in $\log T_{\rm eff}$, $\log g$, and $\rm [Fe/H]$, which are equal to [4750\,K, 
6750\,K], [3.8, 4.6], and [$-$0.5, 0.5], respectively, but made the grid more
dense, and the estimated atmospheric parameters, more certain in comparison with the
results of Paper I.
We show the new FRESCO grid in Fig.\ 1, where we plot the FRESCO reference stars
from Table 8 of Paper I with open circles, the new reference stars, with dots, and 
the eight program stars which fall into the limits of the FRESCO grid, with asterisks.

In Tables 3 and 4, we list the $T_{\rm{eff}}$, $\log g$, and $\rm [Fe/H]$ of 
the 23 SATS and the 10 remaining program stars obtained by the method 
described above. We use a regular font for the atmospheric parameters obtained with
the ELODIE grid and italics for the values obtained with the FRESCO grid.
In the last but one column of the tables, we give the MK spectral 
type assigned using the spectral classification of the reference star which had 
its atmospheric parameters closest to the values computed for the program star. 
We find our classification consistent with the relations between $T_{\rm 
{eff}}$ and spectral type given by Johnson (1966) for dwarfs and giants. In the 
last columns, we list spectral types taken from the Simbad database.

\begin{figure}[htb]
\includegraphics{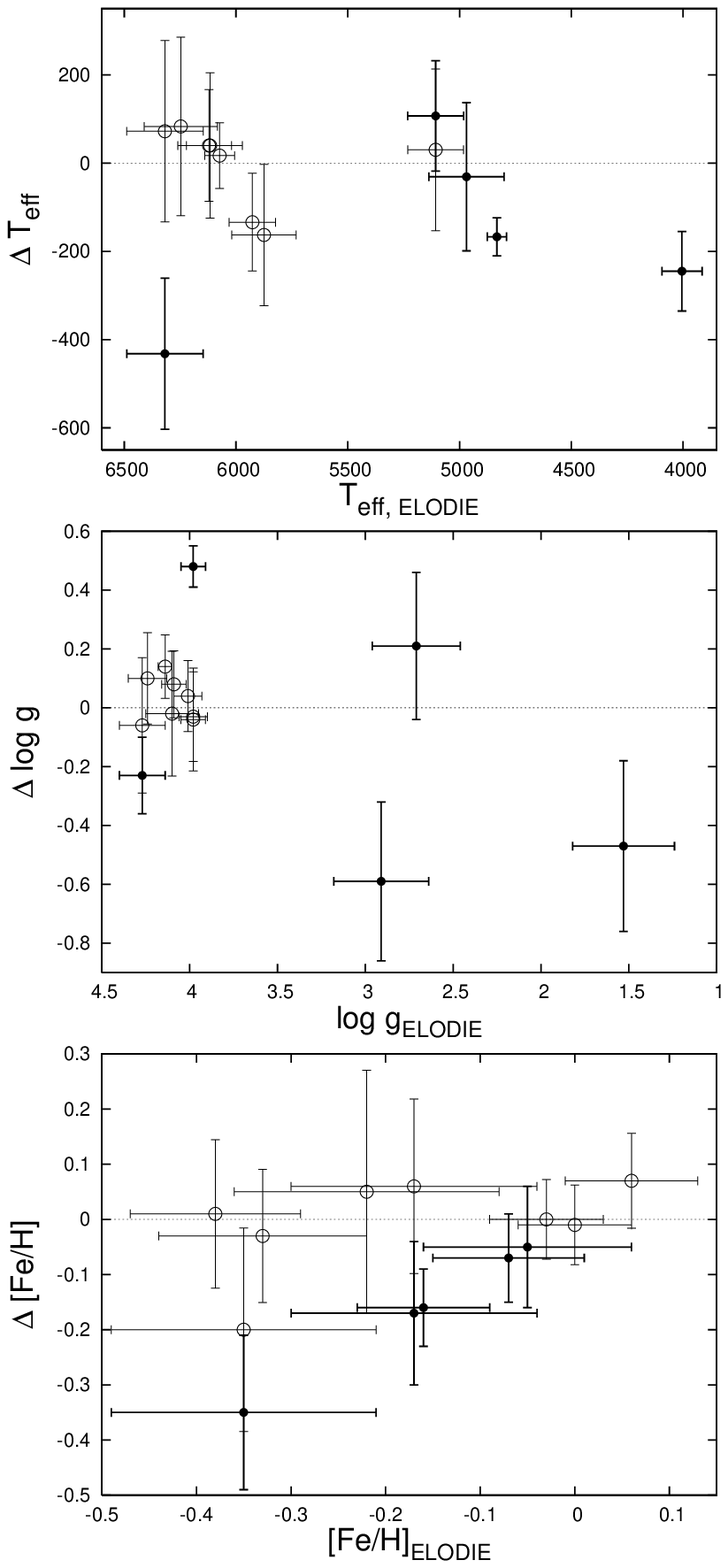}
\FigCap{The differences between $T_{\rm eff}$ (upper panel), $\log g$ (middle panel),
and $\rm [Fe/H]$ (lower panel) determined from the ELODIE and the FRESCO grid
(open circles), or from the ELODIE and the model atmospheres (dots). The dotted
lines indicate zero differences.}
\end{figure}

In Fig.\ 2, we use open circles to plot the difference between the ELODIE 
and FRESCO based values of the effective 
temperature, $\Delta T_{\rm eff}\ = T_{\rm eff}\ ({\rm ELODIE}) - T_{\rm eff}\ ({\rm 
FRESCO})$, the gravity, $\Delta \log g\ = \log g\ ({\rm ELODIE}) - \log g\ ({\rm FRESCO})\, ,$ 
and the metalicity, $\Delta \rm [Fe/H] = \rm [Fe/H] ({\rm ELODIE}) - \rm [Fe/H] ({\rm 
FRESCO})$, for the eight stars for which the atmospheric parameters were determined from both
grids. The weighted mean differences between the atmospheric parameters determined from 
these two grids are equal to $-11\pm$\,30\,K, 0.05$\pm$\,0.03\,dex, and 
0.00$\pm$\,0.03\,dex for $T_{\rm{eff}}$, $\log g$, and $\rm [Fe/H]$, respectively.
This shows that, as in Paper I, the results obtained from both grids agree well and can 
be safely used for our purpose. 

\MakeTable{lrrrrrrll}{12.5cm}{Astrophysical parameters and the MK types of 23 SATS
determined with the use of the ELODIE library (regular font),
FRESCO library (italics) or model atmospheres (bold face). For the
model atmospheres, the solar metalicity was assumed. In
the last column, we list spectral types from the literature}
{\hline\noalign{\smallskip}
HIP   & $T_{\rm eff}$ & s.d. & $\log g$ & s.d. & $\rm [Fe/H]$ &s.d. & MK &Lit.\\
\noalign{\smallskip}\hline\noalign{\smallskip}
 91841  & 4833 &  43 & 2.91 & 0.27 & -0.07 & 0.08 & G9III  &K0\\
        &\bf 5000&   &\bf 3.5 &    &\bf 0.0\\
 92132  & 4947 & 155 & 2.50 & 0.16 & -0.57 & 0.10 & G3IV   &G5\\
 92775  & 5915 & 182 & 4.06 & 0.29 & -1.68 & 0.19 & sdF8   &G2\\
 92961  & 5927 & 104 & 4.24 & 0.11 &  0.00 & 0.06 & G0V    &F8\\
        &\it 6061&\it  38&\it 4.14&\it 0.11&\it 0.01 &\it 0.04 \\
 93320  & 5107 & 125 & 4.27 & 0.13 & -0.35 & 0.14 & K1V    &K3\\
        &\it 5077&\it 134&\it 4.33&\it 0.19&\it -0.15&\it 0.12 \\
        &\bf 5000&   &\bf 4.5 &    &\bf 0.0\\
 93469  & 6318 & 171 & 3.98 & 0.07 & -0.17 & 0.13 & F8IV   &G0\\
        &\it 6246&\it 114&\it 4.02&\it 0.16&\it -0.23&\it 0.09 \\
        &\bf 6750&   &\bf 3.5 &    &\bf 0.0\\
 93607  & 6121 & 101 & 4.09 & 0.07 & -0.03 & 0.06 & F6IV   &F5\\
        &\it 6081&\it  76&\it 4.01&\it 0.09&\it-0.03&\it 0.04 \\ 
 93687  & 4865 &  78 & 2.66 & 0.09 & -0.07 & 0.12 & G9III  &K0\\
 93879  & 6116 & 144 & 3.98 & 0.08 & -0.38 & 0.09 & F8IV   &F8\\
        &\it 6076&\it  80&\it 4.01&\it 0.13&\it -0.39&\it 0.10 \\
 94022  & 4005 &  90 & 1.53 & 0.29 & -0.16 & 0.07 & K4III  &K0\\
        &\bf 4250&   &\bf 2.0 &    &\bf 0.0\\
 94292  & 4895 & 152 & 2.68 & 0.32 & -0.27 & 0.21 & G8III  &G8V\\
 94409  & 4011 &  90 & 1.63 & 0.20 & -0.13 & 0.05 & K4III  &M0\\
 94918  & 5875 & 144 & 4.10 & 0.15 & -0.22 & 0.14 & G0V    &G2V\\
        &\it 6038&\it  70&\it 4.12&\it 0.15&\it -0.27&\it 0.17 \\
 94952  & 4072 &  66 & 1.75 & 0.10 & -0.19 & 0.06 & K4III  &K5\\
 94976  & 4976 & 156 & 2.55 & 0.09 & -0.24 & 0.21 & G5III  &K0V\\
 95237  & 4892 &  97 & 2.63 & 0.08 & -0.11 & 0.12 & G7III  &G5\\
 95654  & 6284 & 173 & 3.97 & 0.09 & -0.08 & 0.08 & F6IV   &F5\\
 96751  & 4601 &  44 & 2.30 & 0.16 & -0.11 & 0.04 & K1III  &K2\\
 96846  & 4599 &  48 & 2.27 & 0.12 & -0.07 & 0.05 & K1III  &K2\\
 97247  & 4040 &  72 & 1.74 & 0.07 & -0.17 & 0.08 & K4III  &K5\\
 97439  & 5370 & 118 & 1.49 & 0.34 &  0.06 & 0.07 & G2Ib   &G2\\
 97513  & 4159 & 152 & 1.72 & 0.33 & -0.20 & 0.06 & K3III  &K0\\
 98793  & 4969 & 168 & 2.71 & 0.25 & -0.05 & 0.11 & G8III  &--\\
       &\bf 5000&   &\bf 2.5 &    &\bf 0.0 \\
\noalign{\smallskip}\hline
}

\MakeTable{lrrrrrrll}{12.5cm}{Astrophysical parameters and MK spectral types of 10
remaining stars determined with the use of the ELODIE library (regular
fonts) or FRESCO library (italics). In
the last column, we list spectral types from the literature}
{\hline\noalign{\smallskip}
HIP   & $T_{\rm eff}$ & s.d. & $\log g$ & s.d. & $\rm [Fe/H]$ &s.d. & MK &lit.\\
\noalign{\smallskip}\hline\noalign{\smallskip}
 92053  & 4013 &  90 & 1.71 & 0.10 & -0.16 & 0.07 & K4III  &M0\\
 92941  & 4866 &  58 & 2.69 & 0.06 &  0.02 & 0.08 & G9III  &K0\\
 93755  & 4673 & 149 & 2.72 & 0.15 & -0.41 & 0.12 & K0III  &K0\\
 94967  & 4912 & 176 & 2.64 & 0.32 & -0.47 & 0.12 & G8IV   &G8V:\\
 95859  & 6073 &  67 & 4.14 & 0.04 &  0.06 & 0.07 & F8V    &F8\\
        &\it 6056&\it  32&\it 4.00&\it 0.10&\it-0.01&\it 0.05 \\
 95913  & 4859 &  56 & 2.79 & 0.15 & -0.16 & 0.10 & G9III  &K0\\
 96642  & 6247 & 164 & 4.01 & 0.08 & -0.33 & 0.11 & F5V    &F5\\
        &\it 6164&\it 118&\it 3.97&\it 0.09&\it -0.30&\it 0.05 \\
 96699  & 5019 & 142 & 2.80 & 0.22 & -0.03 & 0.07 & G8III  &K0\\
 97576  & 4593 &  46 & 2.27 & 0.14 & -0.07 & 0.06 & K1III  &K2\\
 97671  & 4357 & 171 & 2.03 & 0.23 & -0.14 & 0.12 & G9III  &K0\\
\noalign{\smallskip}\hline
}

\subsection{From Model Atmospheres}

For the five SATS observed at the FLWO, we derived global atmospheric parameters using
model atmospheres. As in Paper I, for each program star
we used one-dimensional correlations to identify the 
template in the library of synthetic spectra that gives the best match with the observed 
spectrum, and we chose the template that gave the highest peak correlation value averaged 
over all the observed spectra. We assumed solar metalicity for all 
stars. The parameters are printed with a bold face font in Table 3. 

In Fig.\ 2, we use dots to plot the difference between $T_{\rm eff}$, $\log g$ and $\rm 
[Fe/H]$ obtained from the ELODIE grid and from model atmospheres. As can be seen from
the figure, the ELODIE and model-atmosphere results agree satisfactorily to within
their error bars. The highest discrepancy occurs in case of HIP\,93469 for which the 
model-atmosphere value is 432\,K higher than that obtained from the ELODIE grid.

\section{Projected Rotational Velocity}

In Table 5, we list projected rotational velocities of the program stars together with
their standard deviations. The $v\sin i$ for the stars observed at CAO was determined 
with the Full Width at Half Maximum (FWHM) method for each order of the echelle spectrum 
which did not contain broad spectral lines. As templates, we used a grid of rotationally 
broadened spectra of a non-rotating star having $T_{\rm eff}$, $\log g$, and $\rm [Fe/H]$ 
similar to that of the program star. The upper limit of 5~km\,s$^{-1}$ was estimated 
according to the instrumental resolution of the spectrograms.

For the five SATS observed at the FLWO, we have used the Kurucz model spectra for the 
determination of $v\sin i$. In this method, we compared each observed spectrum with
a library of synthetic spectra using correlation techniques described in Sec.\ 4.2.
A typical standard deviation of these determinations is equal to 1 or 2 km\,s$^{-1}$.
We list these values in Table 5, in the column headed ``model''. 

As can be seen from Table 5, the values of $v\sin i$ obtained from the two separate 
sets of data by means of the above-mentioned two methods agree well. 

\MakeTable{lrrr|lrr}{12.5cm}{Projected rotational velocities determined from a grid of Kurucz 
model spectra and by means of the FWHM method}
{\hline\noalign{\smallskip}
HIP   &$v\sin i$&$v\sin i$&s.d.&HIP   &$v\sin i$&s.d.\\
  &[km\,s$^{-1}$]  &[km\,s$^{-1}$] &[km\,s$^{-1}$]    &   &[km\,s$^{-1}$] &[km\,s$^{-1}$]   \\
  &\tiny (model) &\tiny (FWHM) &\tiny (FWHM)    &   &\tiny (FWHM) & \tiny (FWHM)  \\
\noalign{\smallskip}\hline\noalign{\smallskip}
\multicolumn{4}{c}{SATS}&\multicolumn{3}{c}{The remaining stars}\\
\noalign{\smallskip}\hline\noalign{\smallskip}
91841   &  0.5 & $<$5 &     & 92053   & $<$5        \\
92132   &      & $<$5 &     & 92941   & $<$5        \\
92775   &      &  7.0 & 1.8 & 93755   & $<$5        \\
92961   &      & $<$5 &     & 94967   &  5.2 &  1.6 \\
93320   &  1.5 & $<$5 &     & 95859   & $<$5        \\
93469   & 11.0 &  8.5 & 2.0 & 95913   & $<$5        \\
93607   &      & $<$5 &     & 96642   & 20.5 &  2.0 \\
93687   &      & $<$5 &     & 96699   & $<$5        \\
93879   &      & 15.7 & 1.2 & 97576   & $<$5        \\
94022   &  0.0 & $<$5 &     & 97671   & $<$5        \\
94292   &      & $<$5 &     & \\
94409   &      & $<$5 &     & \\
94918   &      & $<$5 &     & \\
94952   &      & $<$5 &     & \\
94976   &      & $<$5 &     & \\
95237   &      & 17.7 & 0.7 & \\
95654   &      & 12.2 & 1.1 & \\
96751   &      & $<$5 &     & \\
96846   &      & $<$5 &     & \\
97247   &      &  5.3 & 1.7 & \\
97439   &      & 12.3 & 1.6 & \\
97513   &      & $<$5 &     & \\
98793   & 9.0  & $<$5 &     & \\
\noalign{\smallskip}\hline                                    
}

\section{Summary}

We present spectroscopic observations of 23 secondary candidates for Kepler 
asteroseismic targets (SATS) and of 10 other stars. The observations were 
carried out at the {\it M.G.  Fracastoro\/} station of the Catania Astrophysical 
Observatory and the F.L.\ Whipple Observatory, Mount Hopkins, Arizona.  

We find that all SATS but one, HIP\,92775, listed as a subdwarf by Ryan \& Norris
(1991), have solar-like metalicity or are slightly metal-deficient. They range in 
spectral type from early F to late K and therefore we expect all of them to show
solar-like oscillations. 17 SATS are classified in this paper as subgiants or giants. 
These stars are particularly interesting from the asteroseismic 
point of view because the predicted amplitude of solar-like oscillations 
in the evolved stars is expected to be higher than in dwarfs (see, e.g., 
Kjeldsen \& Bedding~1995). 

Our sample contains also HIP\,97439 (= V1154 Cyg), G2Ib, which is a well-known 
Cepheid variable. The light curve of this star has been measured photoelectrically 
in Johnson UBV filters by Wachmann (1976), Szabados (1977) and Henden (1979), 
in UBVR filters by Berdnikov (1987) and Berdnikov (1993), and in uvby$\beta$ filters 
by Arellano et al.~(1998). The typical precision of these observations was $\pm~0.01$ 
mag. The only CCD observations of V1154 Cyg were made in blue light by Turner et 
al.~(1999) and have the precision of around $\pm 0.3$ mag.

V1154 Cyg was included in the search for double mode Cepheids carried out by Szabados 
(1977) and Henden (1979), but was not classified as such. We note, 
however, that the relatively short period of pulsations of V1154 Cyg, equal to 4.925537 
days (Samus et al.~2004), makes this star similar to the 11 double mode Cepheids listed
by Balona (1985) and to the four Cepheids which were supposed by Kovtyukh et al.~(2003) 
to show non-radial pulsations in line profiles. We expect that the 
high-precision Kepler photometry will allow checking whether V1154 Cyg indeed does 
not show any of the above mentioned phenomena, or they are present but with an 
amplitude too small to allow their detection in the photometric data available 
to date.

From all spectrograms, we derive the radial velocities. The results are given in Table 
1, available in electronic form from the Acta Astronomica Archive (see the cover page). 
In addition, the spectrograms obtained at the {\it M.G.  Fracastoro\/} station of the Catania 
Astrophysical Observatory are used to determine the effective temperature, surface 
gravity, metalicity, and MK type by means of the ROTFIT code (Frasca et al. 2003, 2006). 
The spectrograms obtained at the F.L.\ Whipple Observatory are used to determine the 
effective temperature and surface gravity by means of a two-dimensional correlation technique 
TODCOR (Zucker \& Mazeh~1994 and Torres et al.~2002). The results obtained from these two 
different methods applied to two different data sets, given in Table 3, agree well 
in all but one case. We also estimate the projected rotational velocity from two separate sets of 
data using two independent methods (see Table 5) obtaining good agreement.

\Acknow{This work was supported by MNiSW grant N203 014 31/2650, the University of 
Wroc{\l}aw grants 2646/W/IA/06 and 2793/W/IA/07, the Italian government 
fellowship BWM-III-87-W{\l}ochy/ED-W/06 and the Socrates-Erasmus 
Program ``Akcja 2'' 2006-2007, contract No.\ 33. 

J.M.-\.Z.\ acknowledges the European Helio-- and Asteroseismology Network HELAS for the
financial support and thanks Marcin Gie{\l }da for help in observations.

We acknowledge the partial support from the Kepler mission under 
cooperation agreement NCC2-1390 (D.W.L., PI).}

\end{document}